\documentclass[sigconf]{acmart}
\AtBeginDocument{%
  }

\usepackage{algorithmic}
\usepackage{graphicx}
\usepackage{textcomp}
\usepackage{xcolor}
\usepackage{algorithm}

\usepackage{multirow}

\setcopyright{none}
\renewcommand\footnotetextcopyrightpermission[1]{}
\begin{document}


\title{QuSplit: Achieving Both High Fidelity and Throughput via Job Splitting on Noisy Quantum Computers}

\author{Jinyang Li}
\affiliation{%
  \institution{George Mason University}
  \department{Department of Electrical and Computer Engineering}
  \city{VA}
  \country{USA}
}
\email{jli56@gmu.edu}

\author{Yuhong Song}
\affiliation{%
  \institution{George Mason University}
  \department{Department of Electrical and Computer Engineering}
  \city{VA}
  \country{USA}
}

\author{Yipei Liu}
\affiliation{%
  \institution{George Mason University}
  \department{Department of Electrical and Computer Engineering}
  \city{VA}
  \country{USA}
}

\author{Jianli Pan}
\affiliation{%
  \institution{George Mason University}
  \department{Department of Information Sciences and Technology}
  \city{VA}
  \country{USA}
}

\author{Lei Yang}
\affiliation{%
  \institution{George Mason University}
  \department{Department of Information Sciences and Technology}
  \city{VA}
  \country{USA}
}

\author{Travis Humble}
\affiliation{%
  \institution{Oak Ridge National Laboratory}
  \city{TN}
  \country{USA}
}

\author{Weiwen Jiang}
\affiliation{%
  \institution{George Mason University}
  \department{Department of Electrical and Computer Engineering}
  \city{VA}
  \country{USA}
}
\email{wjiang8@gmu.edu}



\begin{abstract}
With the progression into the quantum utility era, computing is shifting toward quantum-centric architectures, where multiple quantum processors collaborate with classical computing resources. Platforms such as IBM Quantum and Amazon Braket exemplify this trend, enabling access to diverse quantum backends. However, efficient resource management remains a challenge, as quantum processors are highly susceptible to noise, which significantly impacts computation fidelity. Additionally, the heterogeneous noise characteristics across different processors add further complexity to scheduling and resource allocation.
Existing scheduling strategies typically focus on mapping and scheduling jobs to these heterogeneous backends, which leads to some jobs suffering extremely low fidelity. Targeting quantum optimization jobs (e.g., VQC, VQE, QAOA) — among the most promising quantum applications in the NISQ era — we hypothesize that executing the later stages of a job on a high-fidelity quantum processor can significantly improve overall fidelity. To verify this, we use VQE as a case study and develop a Genetic Algorithm-based scheduling framework that incorporates job splitting to optimize fidelity and throughput. 
Experimental results demonstrate that our approach consistently maintains high fidelity across all jobs while significantly enhancing system throughput. Furthermore, the proposed algorithm exhibits excellent scalability in handling an increasing number of quantum processors and larger workloads, making it a robust and practical solution for emerging quantum computing platforms. To further substantiate its effectiveness, we conduct experiments on a real quantum processor, IBM Strasbourg, which confirm that job splitting improves fidelity and reduces the number of iterations required for convergence.
\end{abstract}

\maketitle
\settopmatter{printacmref=false}
\setcopyright{none}
\renewcommand\footnotetextcopyrightpermission[1]{}
\pagestyle{plain}

\section{Introduction}

Quantum computing holds great promise to revolutionize domain applications, such as computational chemistry \cite{motta2022emerging, von2018quantum,levine2009quantum, shao2006advances,feniou2023overlap,cao2019quantum, peruzzo2014variational,izmaylov2019unitary,izsak2023quantum} and material science \cite{bauer2020quantum,alexeev2024quantum, teale2022dft,bochevarov2013jaguar, hafner2007materials, ebbesen2016hybrid,reiser2022graph,westermayr2021perspective}, and optimizations \cite{jiang2021co, wang2021exploration, jiang2021machine, stein2022quclassi, jing2022rgb, zeng2022multi, tacchino2020quantum, yan2020nonlinear, gratsea2022exploring}, outperforming state-of-the-art classical computing significantly in terms of speed, accuracy, or the ability to handle larger problem sizes.
Along with the emergence of quantum computers, the computing
paradigm gradually shifts toward quantum-centric computing, where multiple quantum processors collaborate with classical computers.
These quantum processors are inherently heterogeneous due to the presence of different quantum bit (qubit) technologies, such as superconducting qubits (e.g., IBM, Regitte), trapped ion qubits (e.g., IonQ, Quantinuum) and neutral atoms (e.g., QuEra, Infleqtion), bringing quite varied noise levels.
Even for platforms with the same qubit technology, say IBM Quantum, the noise levels are different from processor to processor.
The emergence of quantum-centric computing and the presence of heterogeneity call for automated resource management with two fundamental needs: (1) high quantum processor utilization or job execution throughput and (2) high execution fidelity across quantum processors.




As we are still in the infancy of quantum-centric computing, most vendors require users to specify quantum processors for their jobs, which manually significantly impedes domain scientists who lack quantum knowledge for quantum-augmented scientific discovery.
Only until recently, there emerging works for automated parallel scheduling algorithm \cite{Orenstein2024qgroup,liu2024qucloud}; however, these quantum job scheduling methods follow the traditional scheduling problem developed for classical computing, that is, they focus on assigning one entire job to a specific backend to satisfy the qubit constraints, minimizing makespans, and maximizing fidelity.
Compared with the classical scheduling problem, the new additional objective on fidelity brings great challenges.
Firstly, it creates the conflict to minimize makespans since every job desires the quantum processor with the lowest noise level for high fidelity.
More importantly, secondly, it has a great pitfall if the optimizer uses average fidelity for optimization, which can easily lead to unfair treatment on different jobs, but a soundness average fidelity.

In response to these challenges, we believe application-specific design would be a great help.
To this end, we first investigate the most promising and popular quantum applications,
and it is concluded that the variational quantum algorithms, including VQC, VQE, and QAOA, are among the most promising approaches in the NISQ era. 
Then, we rethought whether job-level granularity is the best practice for scheduling, and we found that job split can be extremely useful.
By placing the front stage of the job to the higher noisy quantum processors and the tail stage to the less noisy quantum processors, one can achieve high fidelity while fully utilizing the heterogeneous quantum processors.






Based on such an inspirational observation, we propose \textit{QuSplit}, representing a novel framework that introduces job-splitting into the quantum scheduling process. QuSplit assigns different stages of a quantum job to backends with varying noise levels, enabling high-fidelity execution for all jobs while improving overall throughput. This approach not only enhances performance but also optimizes resource utilization in quantum-centric computing.

The main contributions of this paper are summarized as follows:
\begin{itemize}
    \item We propose \textit{QuSplit}, a novel job-splitting framework for quantum job scheduling.
    \item We devise an efficient genetic algorithm for job scheduling with the best tradeoff between fidelity and throughput online.
    \item We validate QuSplit through extensive experiments using real noise models, demonstrating its superiority over naive scheduling baseline methods.
\end{itemize}

The remainder of this paper is organized as follows: Section II discusses the background and related work. Section III presents our observations and motivation. Section IV introduces the QuSplit methodology. Section V details the experimental results. Finally, Section VI concludes the paper and discusses future work.

\section{Background and Related Work}

\subsection{Quantum-centric computing}

Over the past decade, quantum computing has achieved tremendous advances in both hardware and software, including continuously increasing quantum bit (qubit) numbers and emerging applications from different domains.
For the foreseeable future, a highly coupled and integrated infrastructure with quantum computers and classical computers will be necessary to ensure efficient and large-scale 
scientific research and innovation, known as Quantum-Centric Computing, which has emerged already, evidenced by IBM Quantum with 14 quantum processors \cite{ibmeurope} and Amazon Braket \cite{amazonbraket} with 7 quantum computers from IonQ, IQM, QuEra, and Rigetti.

\subsection{Quantum Jobs and Applications}

In today's Noisy Intermediate-Scale Quantum (NISQ) era, variational quantum algorithms (VQA) are regarded as one of the most promising quantum algorithms, offering potential quantum advantages in domains like optimization and chemistry.
The VQA, such as Variational Quantum Eigensolver (VQE) and Quantum Approximate Optimization Algorithm (QAOA), leverage a quantum-classical hybrid approach that allows quantum computers to focus on tasks where they have an advantage.
For example, VQE is widely used to solve problems with quantum chemistry. It utilizes a parameterized quantum circuit (ansatz) to approximate the ground state of a Hamiltonian. A classical optimizer iteratively updates the circuit parameters to minimize the expectation value of the Hamiltonian, thereby converging to the lowest energy state.

The performance of variational quantum algorithms is highly sensitive to backend noise levels. In VQE, quantum noise affects the accuracy of expectation value calculations and can cause deviations from the true ground state energy. As such, the choice of backend plays a crucial role in determining the fidelity of VQE results. 
As VQE is considered a representative of varied variational quantum algorithms, we use VQEs for different molecules as the target applications to evaluate the effectiveness of our proposed job-splitting method in this work.

\subsection{Quantum Scheduling}
In quantum-centric computing with multiple heterogeneous quantum processors, efficient scheduling methods for various quantum applications are essential to optimize the utilization of available quantum processors, especially as noise levels vary across different quantum devices. Several works have explored quantum scheduling strategies \cite{ravi2021adaptive1,romero2024quantum,zhang2022research} aimed at optimizing backend assignments to improve fidelity or minimize execution time. 
Most recently, \cite{Orenstein2024qgroup,liu2024qucloud} developed parallel scheduling algorithms.
These methods typically focus on assigning entire quantum jobs to specific backends based on their characteristics, such as noise levels and qubit connectivity. 
Some other approaches also involve partitioning a single backend's qubits to run multiple jobs concurrently for resource sharing, maximizing resource utilization \cite{liu2021qucloud,niu2023enabling}.

However, as one entire job is always mapped to one quantum processor for execution, there exist jobs suffering from significant fidelity sacrifice due to the mapped quantum processor having high noise levels. In this case, maximizing average fidelity will lead to unfairness in scheduled jobs. 

Unlike all these works, this work explores the potential of improving the fidelity of each individual job via 
splitting a single quantum job into multiple stages.
From our motivation examples, by strategically assigning different stages of jobs to quantum computers, it is possible to simultaneously boost throughput and improve fidelity.
The following section presents our observations and explains how they light our approach to optimizing quantum job scheduling.




\section{Observation and Motivation}


This section presents our observations from preliminary experiments, which highlight the opportunities in quantum job scheduling via job splitting. These observations motivate our proposed QuSplit framework. The experiments in this section are conducted using VQE to solve the $H_2$ molecular with a 2-qubit ansatz.

\subsection{Observation 1: Different noise levels lead to different performance}

\begin{table}[t]
\centering
\caption{Comparison of Backend Properties and VQE Results for the energy estimation of $H_2$ Molecular.}
\label{tab:observation1}
\tabcolsep 2.8 pt
\footnotesize
\renewcommand{\arraystretch}{2}
\begin{tabular}{|c|ccc|c|cc|}
\hline
\multirow{2}{*}{Backends} & \multicolumn{3}{c|}{Backend Properties} & \multicolumn{3}{c|}{VQE for $H_2$ molecular} \\ \cline{2-7}
 & \multicolumn{1}{c|}{1-q Error} & \multicolumn{1}{c|}{2-q Error} & Read Error & \multicolumn{1}{c|}{Ref. Value} & \multicolumn{1}{c|}{Approx. Energy}  & Error \\ 
\hline
ibm\_nazca & \multicolumn{1}{c|}{3.38E-04} & \multicolumn{1}{c|}{3.12E-02} & 2.35E-02 & \multirow{2}{*}{-1.86} & \multicolumn{1}{c|}{-1.37}  & 0.49 \\ \cline{1-4}
\cline{6-7}
IonQ & \multicolumn{1}{c|}{3.00E-04} & \multicolumn{1}{c|}{2.12E-03} & 5.10E-03 &  & \multicolumn{1}{c|}{-1.79} & 0.07 \\ \hline
\end{tabular}
\end{table}

Table \ref{tab:observation1} presents the VQE results for the $H_2$ molecular on two different backends: ibm\_nazca and IonQ. Each backend exhibits unique noise characteristics, as reflected in their respective 1-qubit, 2-qubit, and readout errors on average in Table \ref{tab:observation1}. The IonQ backend, with significantly lower error rates (e.g., 5.10E-03 v.s. 3.12E-02 in readout error), achieves an approximate energy of -1.79, which is much closer to the reference value of -1.86, resulting in a smaller error of 0.07. In contrast, the ibm\_nazca backend, with higher error rates, converges to an approximate energy of -1.37, leading to a larger error of 0.49. This demonstrates how backend noise impacts the fidelity of VQE results.

This observation reveals a strong correlation between backend noise and VQE performance. It underscores the need for strategies that intelligently leverage high-fidelity backends to achieve better performance.

\begin{figure}[t]
    \centering
    \includegraphics[width=0.48\textwidth]{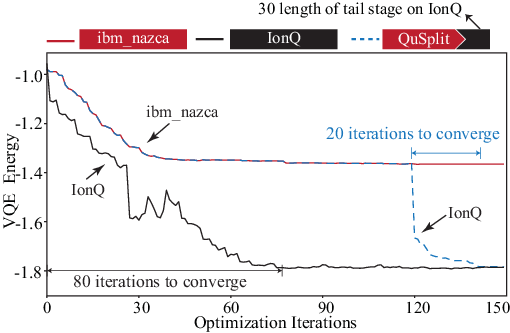}    
    \caption{Impact of job splitting on VQE energy and convergence efficiency.}
    \label{fig:observation2}
\end{figure}

\begin{figure}[t]
    \centering
    \includegraphics[width=0.48\textwidth]{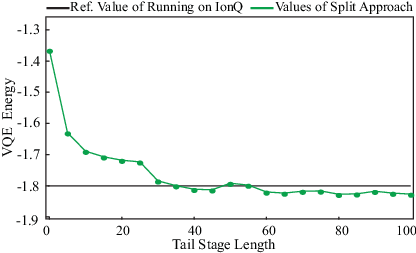}    
    \caption{Effect of tail stage length on VQE energy with job splitting method.}
    \label{fig:observation3}
\end{figure}
\subsection{Observation 2: Splitting jobs can help improve performance}
Figure \ref{fig:observation2} illustrates the benefits of job splitting for the same VQE job by comparing different execution strategies across backends. The ``(green) Split" curve shows the performance when the VQE job begins on a high-noise backend (ibm\_nazca) and switches to a low-noise backend (IonQ) for the tail stage (30 iteration length in our example). This split approach achieves convergence within 20 iterations on IonQ, compared to 80 iterations required if the job is run entirely on IonQ.

This observation demonstrates two key insights: First, 
the length reduction from 80 iterations to 20 iterations reflects the usefulness of running the initial stage on a noisier backend. The progress on the noisier backend can contribute to the optimization process when the job is switched to a better backend.
Second, compared with running the entire job on the noisier backend, splitting the job and executing the tail stages at better backends can significantly improve performance.
As such, quantum jobs can achieve high fidelity while fully utilizing the noisier backends.




\subsection{Observation 3: Split ratio impacts the performance improvement}
Now, an interesting question is how many iterations are needed to be executed on the backend with a lower noise level.
We perform an exploration on 
the impact of various numbers of iterations allocated to the low-noise backend (IonQ) during the tail stage of a VQE job. 
Results are reported in Figure \ref{fig:observation3}, where the x-axis represents the tail stage length (i.e., the number of iterations executed on IonQ), and the y-axis shows the corresponding VQE energy. 
This figure shows that the VQE energy improves significantly as the tail stage length increases, eventually approaching the reference value.
However, after a certain number, executing more iterations on less noisy backends will not further improve the performance.
This observation emphasizes the opportunities to make tradeoffs between application performance and system throughput.


\subsection{Motivation}

In summary, the observations mentioned above provide several insights into designing a high-quality quantum job scheduling:
\begin{itemize}
    \item The significant performance gap between heterogeneous backends necessitates careful job allocation.
    \item Job-splitting offers a promising way to improve performance if the scheduler can strategically utilize less noisy backends during tail stages.
    \item The split ratio must be carefully examined to balance performance improvement and resource utilization.
\end{itemize}

These insights inspire the development of a novel framework, which can dynamically determine the optimal job-splitting strategies to achieve high performance and efficient resource utilization for a large set of quantum jobs.
The following sections detail the methodology and present experimental results demonstrating its advantages over quantum scheduling without job splitting.

\section{QuSplit Framework}

\subsection{Design Philosophy of QuSplit}
QuSplit introduces a novel job-splitting strategy to address the dual challenges of maintaining high fidelity and maximizing resource utilization in quantum job scheduling. The core idea is to split a quantum job into two stages:
\begin{itemize}
    \item Stage 1: The job runs on a noisier backend (i.e., higher noise level) to take advantage of its availability.
    \item Stage 2: The job switches to a less noisy backend (i.e., lower noise level), improving the fidelity of the final results.
\end{itemize}

This approach allows us to balance throughput and performance by leveraging multiple heterogeneous available quantum backends with varying noise levels.

\subsection{Overview of QuSplit}
\begin{figure}[t]
    \centering
    \includegraphics[width=0.48\textwidth]{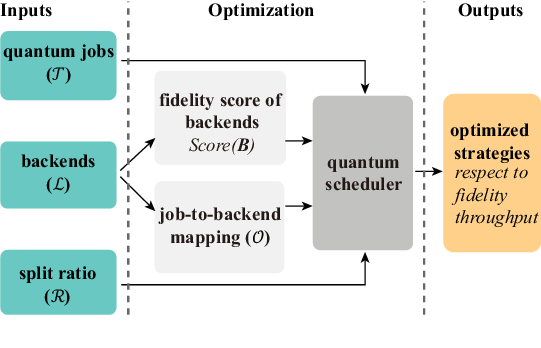}    
    \caption{Overview of the proposed QuSplit framework.}
    \label{fig:overview}
\end{figure}

The QuSplit framework follows a systematic process for efficiently scheduling quantum jobs. The workflow of QuSplit is illustrated in Figure \ref{fig:overview}, including three processes: (1) receiving inputs, (2) scheduling optimization, and (3) generating outputs.


\textbf{Input:}
The input consists of three main components:
\begin{itemize}
    \item $\mathcal{J}$: A list of quantum jobs (i.e., VQE jobs) to be scheduled on multiple quantum processors.
    \item $\mathcal{L}$: The list of available quantum backends.
    \item $\mathcal{R}$: A predefined job split ratios that determine the fraction of iterations assigned to Stage 2.
\end{itemize}

\textbf{Optimization process:}
As shown in Figure \ref{fig:overview}, the optimization process has three components: \textit{i)} fidelity scores of backends; \textit{ii)} a set of job-to-backend mappings; \textit{iii)} a centralized scheduler.


First, we need a metric to judge the quality of backends in $L$.
Such a metric is preferable to be obtained offline to avoid heavy online overhead.
To this end, we develop a proxy approach by executing a set of random benchmark circuits on backends in $L$ offline.
For each benchmark job $j$, we know its reference value $ref_j$.
After one execution on a noisy backend $B$, we can get its real value $real_{j,B}$; then, we can get the deviation $dev_{j,B}=|real_{j,B}-ref_j|$.
Based on these definitions, we define a normalized performance score $score(j,B)$ as below:

\begin{equation}
    score(j,B) = \frac{|ref_j|}{|ref_j| + |dev_{j,B}|}
\end{equation}

To obtain the proxy score of a backend $B$, we will test multiple benchmarking jobs on $B$, denoted as $J$.
Then, the average proxy score will be used for the fidelity metrics for backend $B$ as below:
\begin{equation}\label{eq:scoreb}
score(B)=\frac{\sum_{j\in \mathcal{J}}score(j,B)}{|\mathcal{J}|}
\end{equation}
Kindly note that this score definition can also be used to evaluate the fidelity of VQE jobs executed on quantum backends, where $dev$ represents the deviation between the estimated energy ($real$) and reference energy ($ref$).


Second, in addition to the fidelity score, the scheduler will also need a set of possible job-to-backend mappings, denoted as $\mathcal{O}$, which provide the execution strategies for the job.
Each entire job can be mapped to one backend, and a split job will be mapped to two backends.
There are $|L|+C_{|L|}^{2}=|L|+\frac{|L|\times|L-1|}{2}$ options in total.
For example, suppose we have three backends, say $B1$, $B2$, $B3$, and their score rank is $score(B1)<score(B2)<score(B3)$, we will have the list of possible job-to-backend mappings: [$B1$, $B2$, $B3$, $split_{B1B2}$, $split_{B1B3}$, $split_{B2B3}$] For a job $j$, the selection of $B_1$ indicates it will be entirely mapped to $B_1$ for execution, while 
$split_{B1B2}$ means that the stage 1 of this job will be running on Backend 1, while the stage 2 of it will run on the Backend 2.

In the last step of the optimization process, the quantum scheduler will use the input data, fidelity score, and job-to-backbend mapping set to explore and evaluate various scheduling strategies. 
We will introduce the scheduler in the next subsection.


\textbf{Outputs:}
The output of the optimization process is an optimal scheduling strategy, specifying the backends for each job stage and their execution order with bi-objectives of maximizing the proxy fidelity (scores) and the system throughput.

\subsection{Fidelity-aware Quantum Scheduler}

We formally define the fidelity-aware quantum scheduling, FAQS, as an optimization problem:
Given a list of quantum jobs $\mathcal{J}$, a list of available quantum backends $\mathcal{L}$, yielding to a job-to-backend mapping set $\mathcal{O}$, and the splitting ratios $\mathcal{R}$ of jobs in $\mathcal{J}$, the FAQS problem is to determine:

\begin{itemize}
    \item job mapping $M(\mathcal{J})=\{M(j)\in \mathcal{O}$, for $\forall j\in J$\};
    \item job scheduling strategy $S(\mathcal{J})$;
\end{itemize}
such that both \textbf{throughput} and \textbf{fidelity} can be maximized.
Here, the throughput $TH(Sol)$ of a solution $Sol=\langle M,S\rangle$ under mapping $M$ and scheduling $S$ can be calculated using the number of jobs $|\mathcal{J}|$ and the makespan of $Sol$ (i.e., total execution time, $T$) : $TH(Sol)=\frac{|\mathcal{J}|}{T}$.
On the other hand, fidelity $FI(Sol)$ is defined as the average fidelity score of jobs in $\mathcal{J}$ under the mapping $M$.
For a job $j\in \mathcal{J}$, we use the backend fidelity, defined in Eq. \ref{eq:scoreb}, to approximate its fidelity $FI_j$.
Specifically, if job $j$ is not split, then $FI_j=score(M(j))$. 
Otherwise for a split job $k$, we have $M(k)=split_{B_{p},B_{q}}$, where $B_{q}$ is the less noisy backend.
Then, we will have $FI_k=score(B_{q})$. This is based on the observation that the performance of a split job can approach the performance that the entire job runs on the less noisy backend, i.e., $B_{q}$.
Finally, we have $FI(Sol)=\frac{\sum_{j\in \mathcal{J}}FI_j}{|\mathcal{J}|}$ for the solution $Sol$.




\begin{algorithm}[b]
\caption{Genetic Algorithm for Quantum Job Scheduling}
\label{alg:GA}
\begin{algorithmic}[1]
\STATE \textbf{Input:} Population size $N$, Number of generations $G$, Mutation rate $m$, Elite size $E$
\STATE \textbf{Output:} Optimal scheduling strategy
\STATE Initialize population $P$ with $N$ random strategies
\FOR{generation = 1 to $G$}
    \STATE Evaluate the fitness of each strategy $Sol$ in $P$ using:
    \[
    \text{Fitness}(Sol) = w_1 \cdot TH(Sol) + w_2 \cdot FI(Sol)
    \]
    \STATE Preserve top $E$ elite strategies
    \STATE Select parents from $P$ using roulette wheel selection
    \STATE Apply crossover to produce offspring
    \STATE Apply mutation to offspring with probability $m$
    \STATE Evaluate fitness of offspring
    \STATE Replace worst strategies in $P$ (excluding elites) with offspring
\ENDFOR
\RETURN best strategies $Sol^{*}$ in $P$
\end{algorithmic}
\end{algorithm}

To solve this bi-objective optimization problem, we relax it to solve the following optimization problem:

\begin{equation}\label{eq:sol}
Sol^* = \arg\max_{\forall Sol} \left( w_1 \cdot TH(Sol) + w_2 \cdot FI(Sol) \right)    
\end{equation}
where $w_1$ and $w_2$ are scaling factors to balance throughput and fidelity based on the number of jobs. As more jobs come in, the weight of $w_1$ will be increased to ensure high system throughput. 
At run-time, we apply such a dynamic weight-changing approach to select suitable strategies according to the workload size.

\textbf{Genetic Algorithm (GA):}
To solve the above optimization problem, we apply GA as our optimizer. Kindly note that any other optimizer can be applied in the QuSplit framework.
For the applied GA, it begins by initializing a population of random strategies. Each strategy is evaluated using a fitness function that combines throughput and fidelity, as defined in Eq. \ref{eq:sol}. The fittest strategies are selected as parents, which undergo crossover and mutation to produce new candidate solutions. This process continues for a fixed number of generations or until convergence criteria are met.
The specific steps of the GA used in QuSplit are detailed in Algorithm \ref{alg:GA}.


\subsection{Example of Scheduling Strategy}

\begin{figure}[t]
    \centering
    \includegraphics[width=0.48\textwidth]{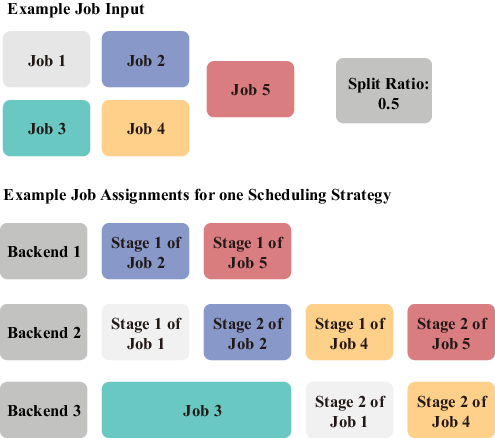}    
    \caption{Example of job assignments for a scheduling strategy.}
    \label{fig:example}
\end{figure}

Figure \ref{fig:example} illustrates how quantum jobs are assigned to different backends under a specific scheduling strategy for 5 jobs with 3 available backends. The strategy used in this example is as follows: 
[$split_{B2B3}$, $split_{B1B2}$, $B3$, $split_{B2B3}$, $split_{B1B2}$]
This strategy specifies how each job is executed, including whether it is split between two backends and which backends are used for each stage.

\textbf{Example Job Input:} 
The top section of the figure shows five quantum jobs (Job 1 to Job 5) and a predefined split ratio of 0.5. This split ratio indicates that for split jobs, 50\% of the iterations will run on the first backend (Stage 1) and the remaining 50\% will run on the second backend (Stage 2).

\textbf{Job Assignments:} 
The bottom section of the figure illustrates how the stages of each job are mapped to the available backends (Backend 1, Backend 2, and Backend 3) according to the specified strategy:
\begin{itemize}
    \item \textbf{Job 1 ($split_{B2B3}$)}: Stage 1 is assigned to Backend 2, and Stage 2 is assigned to Backend 3.
    \item \textbf{Job 2 ($split_{B1B2}$)}: Stage 1 is assigned to Backend 1, and Stage 2 to Backend 2.
    \item \textbf{Job 3 ($B3$)}: The entire job runs on Backend 3.
    \item \textbf{Job 4 ($split_{B2B3}$)}: Stage 1 is assigned to Backend 2, and Stage 2 to Backend 3.
    \item \textbf{Job 5 ($split_{B1B2}$)}: Stage 1 is assigned to Backend 1, and Stage 2 to Backend 2.
\end{itemize}

This example demonstrates the flexibility of QuSplit in distributing job stages across backends to balance performance and resource utilization. 
The split strategy allows leveraging high-fidelity backends for critical computations while utilizing lower-fidelity backends for non-critical stages.

\section{Experimental Results}
\begin{table}[b]
\centering
\caption{Performance Scores of VQE Jobs with Split and Non-Split Configurations Across Different Molecules.}
\label{tab:exp1}
\tabcolsep 5.5 pt
\footnotesize
\renewcommand{\arraystretch}{2}
\begin{tabular}{|c|c|cc|cc|cc|}
\hline
\multirow{2}{*}{} & \multirow{2}{*}{} & \multicolumn{2}{c|}{Non-split} & \multicolumn{2}{c|}{SplitB1\_B3} & \multicolumn{2}{c|}{Improv.} \\ \cline{3-8}
\multirow{-2}{*}{Molecules} & \multirow{-2}{*}{\# Qubits} & \multicolumn{1}{c|}{B1} & B3 & \multicolumn{1}{c|}{$r$ = 0.4} & $r$ = 0.6 & \multicolumn{1}{c|}{$r$ = 0.4} & $r$ = 0.6 \\ \hline
$H_2$ & 2 & \multicolumn{1}{c|}{0.79} & 0.96 & \multicolumn{1}{c|}{0.98} & 0.98 & \multicolumn{1}{c|}{0.19} & 0.19 \\ \hline
$BF$ & 4 & \multicolumn{1}{c|}{0.69} & 0.95 & \multicolumn{1}{c|}{0.78} & 0.86 & \multicolumn{1}{c|}{0.09} & 0.17 \\ \hline
$CO_2$ & 6 & \multicolumn{1}{c|}{0.74} & 0.94 & \multicolumn{1}{c|}{0.88} & 0.90 & \multicolumn{1}{c|}{0.14} & 0.16 \\ \hline
\end{tabular}
\end{table}

\subsection{Experimental Setup}

\textbf{Quantum Jobs:} To evaluate the proposed quantum job scheduling and splitting strategies, we conducted experiments using multiple Variational Quantum Eigensolver (VQE) jobs. Each VQE job corresponds to a quantum chemistry problem, where qubit operators were generated using the \textit{Qiskit Nature} package. The ansatz used for these VQE jobs was \textit{EfficientSU2} with three repetitions, a commonly used variational parameterized circuit for quantum chemistry problems.

\textbf{Backends and Noise Models:} The experiments utilized simulated noise models derived from real quantum backends, labeled as \textbf{B1}, \textbf{B2}, and \textbf{B3}. These noise models were constructed using backend-specific noise information, such as gate error rates, $T_1$, and $T_2$ relaxation times. Noisy simulations were performed using the \textit{Qiskit Aer Estimator}, emulating the behavior of real quantum hardware under realistic noise conditions.

\begin{figure}[b]
    \centering
    \includegraphics[width=0.47\textwidth]{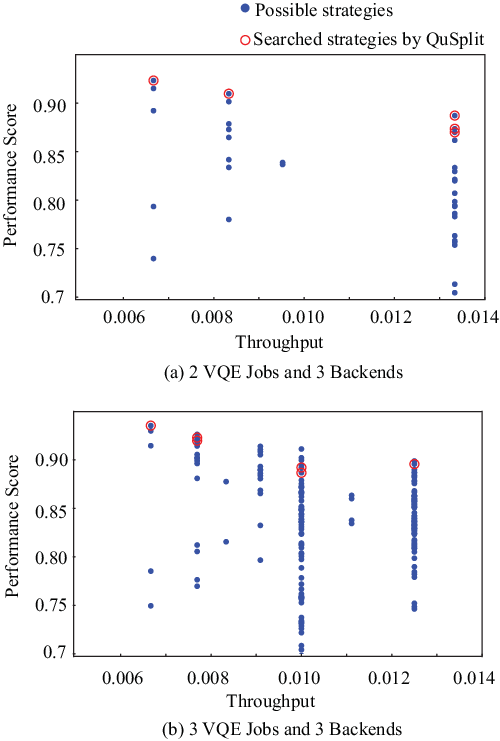}    
    \caption{Design space exploration of job scheduling strategies for VQE jobs using QuSplit.}
    \label{fig:design_space_exp}
\end{figure}

\begin{figure*}[t]
    \centering
    \includegraphics[width=\textwidth]{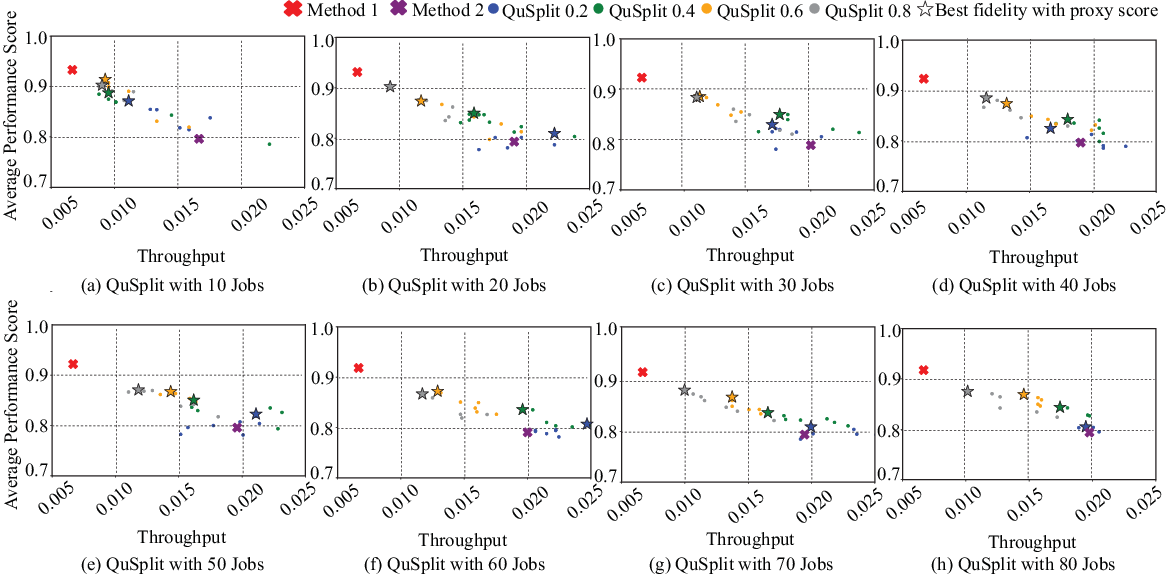}    
    \caption{QuSplit evaluation on large-scale job scheduling scenarios with the split ratio of 0.2, 0.4, 0.6, 0.8 compared with all executions on the best backend (Method 1) and backend round-robin scheduling (Method 2).}
    \label{fig:main_results}
\end{figure*}

\textbf{Splitting Configurations and Job Length:} Each job consisted of 150 iterations, and we evaluated both non-split and split configurations using the following notation:
\begin{itemize}
    \item B1, B2, B3: No split, where the entire job runs on the backend.
    \item splitB1\_B2, splitB1\_B3, etc: Split configurations, where the first stage runs on the first backend (e.g., B1), and the second stage runs on the second backend (e.g., B2 or B3). 
    \item Split ratio $r$: the fraction of iterations allocated to the second stage. ($r$ = 0.2, 0.4, 0.6 or 0.8).
\end{itemize}


\textbf{Optimization Setting:} The optimization of job scheduling strategies was performed using the \textit{Genetic Algorithm (GA)}. The GA operated with a population size of 10, a mutation rate of 0.2, and evolved over 100 generations.


    


\subsection{Evaluation of Job-Splitting}

Table~\ref{tab:exp1} summarizes the performance of VQE jobs for three different molecules ($H_2$, $BF$, and $CO_2$) under multiple scheduling strategies. These include both non-split executions, where the entire job is run on a single backend, and job-splitting configurations (SplitB1\_B3) with different split ratios ($r$ = 0.4 and $r$ = 0.6). Here, B3 represents a higher-fidelity quantum processor, whereas B1 is more error-prone. The results illustrate how job splitting influences performance under varying execution conditions, leading to the following key observations.

\textbf{Performance Improvement through Job-Splitting.} 
Job splitting proves to be a highly effective technique for improving VQE performance compared to executing the entire computation on the noisier backend (B1). For example, in the case of $H_2$, the performance improves from 0.79 (B1) to 0.98 under both split configurations ($r$ = 0.4 and $r$ = 0.6). Similar improvements are seen across the other molecules, suggesting that incorporating a more reliable backend in the latter stages of execution helps mitigate the adverse effects of noise and improve performance.

\textbf{Effect of Split Ratio.}
The proportion of iterations allocated to the high-fidelity backend (B3) plays a crucial role in determining the final performance.
A larger allocation to B3 generally yields better performance. For instance, the VQE performance for the $BF$ molecule achieves a performance score of 0.78 with a split ratio of $r$ = 0.4, which increases to 0.86 with $r$ = 0.6. This pattern reinforces the importance of optimizing the split ratio to strike a balance between leveraging computational efficiency and maximizing result accuracy.

\textbf{Comparison with All B3 Strategy.}
While executing all iterations on B3 generally yields the best performance (e.g., 0.95 for $BF$ and 0.94 for $CO_2$), the split strategy offers a practical alternative. By leveraging the lower-fidelity backend (B1) in the early stages, job-splitting achieves comparable or near-optimal performance. This indicates that job-splitting can be a practical scheduling strategy, particularly when access to high-fidelity quantum processors is constrained.

\subsection{Design Space Exploration}

Figure \ref{fig:design_space_exp} showcases the design space exploration for scheduling strategies involving two VQE jobs (Figure~\ref{fig:design_space_exp}(a)) and three VQE jobs (Figure~\ref{fig:design_space_exp}(b)) across three quantum backends. These experiments represent small-scale job scheduling scenarios, where various execution strategies are evaluated based on their performance score (y-axis) and throughput (x-axis). Each blue dot in the figure corresponds to a unique scheduling strategy, while the strategies identified by QuSplit are highlighted using red circles.

For two VQE jobs, brute-force enumeration of all possible strategies is computationally feasible, allowing the entire design space to be explored. QuSplit effectively identifies high-quality strategies (red circles) that lie on or near the Pareto Frontier, demonstrating its ability to identify optimal trade-offs between performance and throughput. Similarly, when scaling up to three VQE jobs, the design space expands, increasing the number of possible scheduling strategies. Despite this growth, brute-force exploration is still manageable, allowing a full assessment of the landscape. Notably, QuSplit continues to identify near-optimal scheduling configurations, reinforcing its effectiveness in balancing performance and execution efficiency even as the scheduling complexity increases.

\subsection{QuSplit Evaluation on Large-Scale Job Scenarios}

Figure \ref{fig:main_results} illustrates the relationship between throughput and average performance score for various job scheduling strategies, evaluated across different job sizes ranging from 10 to 80 jobs. These experiments represent large-scale job scheduling scenarios, where different methodologies are compared in terms of their ability to balance computational efficiency and execution fidelity. The strategies analyzed include:

\begin{itemize}
    \item \textbf{QuSplit 0.2, 0.4, 0.6, 0.8} (blue, green, orange, and grey points, respectively): Job-splitting strategies generated by the QuSplit method, which distribute jobs across different backends based on specified split ratios.
    \item \textbf{Method 1} (red marker): All jobs are executed exclusively on the highest-fidelity backend (B3), maximizing performance at the cost of reduced throughput.
    \item \textbf{Method 2} (purple marker): Round-robin scheduling \cite{rasmussen2008round,balharith2019round}, which is widely studied in classical and cloud computing, where jobs are cyclically assigned to backends in the sequence of $\langle B3, B2, B1, \dots \rangle$.
\end{itemize}

Based on the above experimental results, we conclude the following key observations:
\begin{itemize}
    \item Our QuSplit can find a favorable balance between throughput and performance, achieving high performance scores while maintaining competitive throughput.
    \item Method 1 consistently delivers the highest performance, as expected, but suffers from severely limited throughput, making it less practical for handling large workloads efficiently.
    \item Method 2 improves throughput by distributing jobs more evenly, yet this comes at the cost of significantly lower performance due to its fixed allocation across backends with varying noise levels.
    \item As the number of jobs increases, QuSplit strategies continue to scale effectively, retaining their advantage in balancing performance and execution efficiency even at 80 jobs.
\end{itemize}

These results highlight the practicality of QuSplit as a scheduling approach for large-scale quantum workloads. By intelligently distributing job execution across backends based on noise-aware split ratios, QuSplit enables efficient resource utilization while preserving computational accuracy.

\subsection{Scheduling Time Analysis}
\begin{figure}[t]
    \centering
    \includegraphics[width=0.48\textwidth]{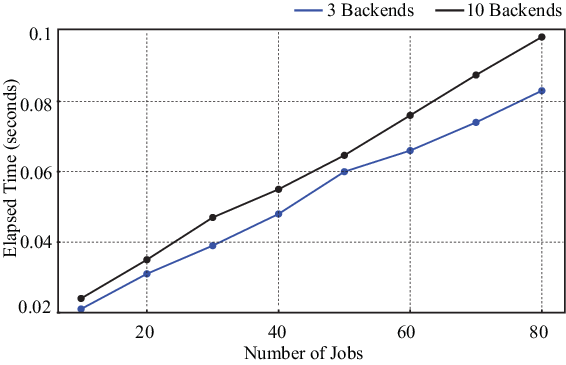}    
    \caption{Elapsed time of QuSplit for different numbers of jobs across 3 and 10 backends.}
    \label{fig:sche_time}
\end{figure}

Figure \ref{fig:sche_time} presents the scheduling time required to allocate jobs across different numbers of backends, comparing configurations with 3 and 10 backends. The x-axis represents the number of jobs, while the y-axis shows the corresponding scheduling time in seconds. 

The results indicate that the scheduling time for 10 backends (black line) is consistently higher than for 3 backends (blue line). This increase is attributed to the growing complexity of searching for optimal job allocations as the number of available backends expands. With more backends, the solution space enlarges, requiring additional computation time to evaluate and determine efficient scheduling strategies.
Despite this complexity, both configurations exhibit a linear increase in scheduling time as the number of jobs grows from 10 to 80. However, even with 10 backends, the scheduling time remains within acceptable limits, demonstrating the scalability of the scheduling algorithm.

\subsection{Validation of Backend Fidelity Score}
\begin{figure}[t]
    \centering
    \includegraphics[width=0.48\textwidth]{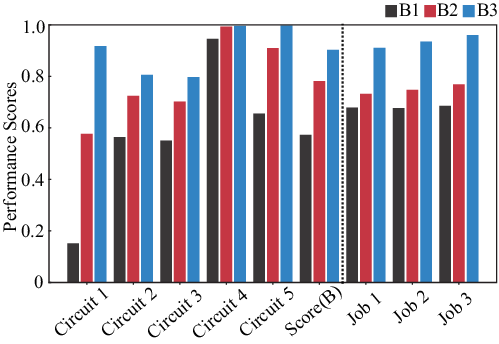}    
    \caption{Validating benchmarks-based backends fidelity score is proportional to the fidelity of quantum VQE jobs.}
    \label{fig:val}
\end{figure}
Figure \ref{fig:val} shows the fidelity scores of five random benchmark circuits and the performance scores of three VQE jobs evaluated on three backends: B1, B2, and B3. Experimental results indicate that backend B3 consistently achieves the highest scores across all circuits, followed by B2 and B1, demonstrating an intuition of lower noise levels correlating with higher performance.
Score(B) metric, shown in Figure~\ref{fig:val}, represents the proxy backend fidelity score computed according to Eq.~\ref{eq:scoreb}. This score serves as a numerical estimation of each backend's expected performance, derived from benchmark circuit results.

For example, Job 1 achieves a score of 0.911 on B3, significantly outperforming its scores on B2 (0.732) and B1 (0.679). Similar patterns are observed for Job 2 and Job 3.

These observations validate the hypothesis that the proxy backend fidelity defined in Eq. \ref{eq:scoreb} is a reliable metric for estimating the potential performance of VQE jobs on a backend. By leveraging these fidelity scores, we can effectively rank quantum backends for job scheduling, ensuring that jobs are directed to the most suitable resources to optimize performance.

\subsection{Distribution of Job Performance Deviation}

\begin{figure}[t]
    \centering
    \includegraphics[width=0.48\textwidth]{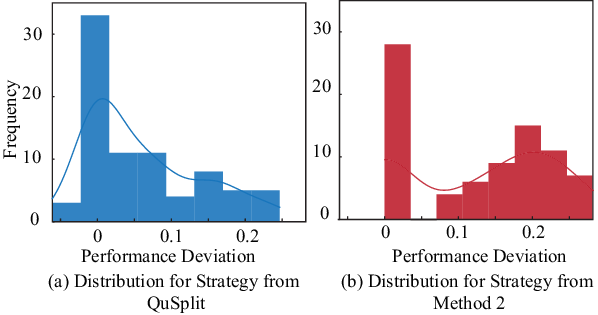}    
    \caption{Distribution of job performance deviation for strategies: (a) result of QuSplit, and (b) baseline Method 2, which exhibits unfairness for jobs executed on heavy noise backend.}
    \label{fig:distribution_gap}
\end{figure}

Figure \ref{fig:distribution_gap} presents the distribution of job performance deviation for two different scheduling strategies: QuSplit and Method 2. The performance deviation for each job is calculated as the difference between its performance on the optimal backend (B3) and the backend assigned in the corresponding strategy.

Figure \ref{fig:distribution_gap} (a) illustrates the performance deviation distribution for jobs scheduled using the QuSplit strategy. The majority of jobs exhibit a small performance deviation close to 0. This highlights the effectiveness of QuSplit in minimizing performance loss.

Figure \ref{fig:distribution_gap} (b) shows the performance deviation distribution for jobs scheduled using Method 2, which follows a round-robin approach. Unlike QuSplit, a significant portion of jobs experience larger performance deviations, even with some exceeding 0.2. This demonstrates that Method 2 fails to effectively leverage high-fidelity devices, resulting in larger deviations from the optimal performance on B3.

This comparison highlights the importance of backend-aware scheduling strategies in minimizing performance degradation and improving the fairness of jobs. QuSplit's ability to prioritize optimal backend selection ensures consistent and efficient job execution, particularly in environments where fidelity significantly impacts performance.
These results emphasize the necessity of adaptive scheduling strategies like QuSplit to achieve robust and high-quality outcomes in heterogeneous computing systems.

\subsection{Real Quantum Computer Results}
\begin{table}[t]
\centering
\caption{Comparison of Backend Properties and VQE Results for the energy estimation of $H_2$ Molecular on real quantum computer.}
\label{tab:real1}
\tabcolsep 2.5 pt
\footnotesize
\renewcommand{\arraystretch}{2}
\begin{tabular}{|c|ccc|ccc|}
\hline
\multirow{2}{*}{Qubit Set} & \multicolumn{3}{c|}{Noise Properties} & \multicolumn{3}{c|}{VQE for $H_2$ molecular} \\ \cline{2-7} 
 & \multicolumn{1}{c|}{1-q Error} & \multicolumn{1}{c|}{2-q Error} & Readout Error & \multicolumn{1}{c|}{Ref. Value} & \multicolumn{1}{c|}{Appro. Energy} & Error \\ \hline
\multirow{2}{*}{{[}60, 61{]}} & \multicolumn{1}{c|}{1.95E-04} & \multicolumn{1}{c|}{\multirow{2}{*}{1.11E-01}} & 7.70E-03 & \multicolumn{1}{c|}{\multirow{2}{*}{-1.86}} & \multicolumn{1}{c|}{\multirow{2}{*}{-1.28}} & \multirow{2}{*}{0.58} \\ \cline{2-2} \cline{4-4}
 & \multicolumn{1}{c|}{9.93E-02} & \multicolumn{1}{c|}{} & 1.02E-01 & \multicolumn{1}{c|}{} & \multicolumn{1}{c|}{} &  \\ \hline
\multirow{2}{*}{{[}99, 100{]}} & \multicolumn{1}{c|}{2.05E-04} & \multicolumn{1}{c|}{\multirow{2}{*}{6.10E-03}} & 2.02E-02 & \multicolumn{1}{c|}{\multirow{2}{*}{-1.86}} & \multicolumn{1}{c|}{\multirow{2}{*}{-1.82}} & \multirow{2}{*}{0.04} \\ \cline{2-2} \cline{4-4}
 & \multicolumn{1}{c|}{5.43E-04} & \multicolumn{1}{c|}{} & 1.07E-02 & \multicolumn{1}{c|}{} & \multicolumn{1}{c|}{} &  \\ \hline
\end{tabular}
\end{table}

\begin{figure}[t]
    \centering
    \includegraphics[width=0.48\textwidth]{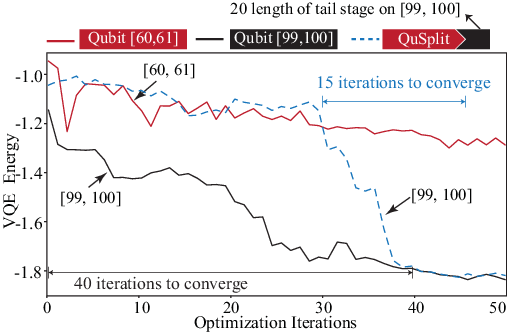}    
    \caption{Impact of job splitting on VQE energy and convergence efficiency on real quantum computer.}
    \label{fig:real2}
\end{figure}

\begin{figure}[t]
    \centering
    \includegraphics[width=0.48\textwidth]{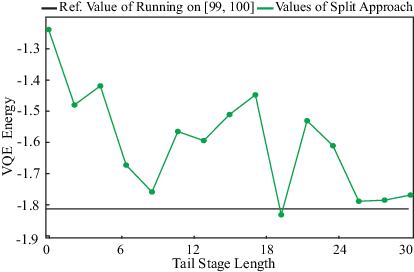}    
    \caption{Effect of tail stage length on VQE energy with job splitting method on real quantum computer.}
    \label{fig:real3}
\end{figure}

To further evaluate the impact of job splitting on VQE performance, we conducted experiments on the IBM quantum computer \textbf{ibm\_strasbourg} using two qubit sets of it: [60,61] and [99,100]. The backend properties and corresponding VQE energy estimations for H$_2$ molecular calculations are summarized in Table~\ref{tab:real1}.

\textbf{Qubit Noise and VQE Accuracy}
Table~\ref{tab:real1} shows the noise characteristics of the two qubit sets. The [99,100] set exhibited lower error rates across single-qubit, two-qubit, and readout operations, whereas the [60,61] set had higher noise. Consequently, the VQE energy computed on [99,100] achieved an approximation of -1.82, closely matching the reference value of -1.86, with a small error of 0.04. In contrast, computations on [60,61] resulted in a larger deviation, with an approximate energy of -1.28 and an error of 0.58.

\textbf{Impact of Job Splitting on Convergence Efficiency}
Figure~\ref{fig:real2} illustrates the optimization curves for VQE execution under different conditions. When the entire optimization is performed on [60,61], convergence requires 40 iterations due to higher noise levels. However, when employing the QuSplit strategy—where the first stage runs on [60,61] and the tail stage transitions to [99,100]—convergence is achieved in just 15 iterations. This demonstrates that utilizing qubits with higher noise for early-stage optimization, followed by a switch to more stable qubits, can significantly reduce the number of required optimization steps.

\textbf{Effect of Tail Stage Length on VQE Accuracy}
Figure~\ref{fig:real3} examines the influence of tail stage length on the final VQE energy. As the number of iterations allocated to the [99,100] tail stage increases, the final energy result improves, approaching the reference value. However, beyond a certain point, additional tail-stage iterations yield diminishing returns, suggesting that an optimal job-splitting ratio might be considered when allocating computational resources.

The results from the \textbf{ibm\_strasbourg} backend reinforce the effectiveness of the QuSplit approach in improving VQE performance. By managing computational workload across qubit sets with varying noise characteristics, we observe improvements in both energy accuracy and convergence efficiency. These findings validate the potential of dynamic job-splitting strategies in practical quantum computing scenarios.

\section{Conclusion}

In this paper, we introduced \textit{QuSplit}, a novel framework for quantum job scheduling that leverages job-splitting to optimize both performance and resource utilization. QuSplit dynamically assigns different stages of quantum jobs to backends with varying noise levels, enabling high-fidelity execution for critical stages while taking full advantage of backends with higher noise levels. Our experiments, conducted with realistic noise models and benchmarked against other scheduling methods, demonstrated the effectiveness of QuSplit in improving average performance scores while balancing resource usage.

One of the key insights from our work is the impact of split ratios on performance. The ability to fine-tune the split ratio allows QuSplit to adapt to different quantum jobs and backend configurations, providing a flexible and efficient scheduling strategy. 
This scalability makes QuSplit particularly well-suited for real-world quantum computing environments, where resource heterogeneity and workload variability are common challenges. 
By effectively balancing performance optimization with the growing demands of quantum workloads, QuSplit is scalable for quantum resource management.

Looking ahead, several promising directions can further enhance the capabilities of QuSplit:

\begin{itemize}
    \item \textbf{Dynamic Prediction of Split Ratios:} One potential improvement is the integration of \textit{machine learning models}, such as transformer-based architectures, to dynamically predict optimal split ratios for quantum jobs. By training on historical data and job characteristics, these models could provide split ratio recommendations tailored to specific quantum jobs and backend conditions.
    
    \item \textbf{Scalability and Generalization:} While this work focused on VQE as a representative quantum application, QuSplit's methodology can be generalized to other quantum algorithms, such as Variational Quantum Circuits (VQC) and Quantum Approximate Optimization Algorithm (QAOA). Future research could explore how job-splitting impacts these algorithms, particularly in larger-scale quantum systems with more complex workloads.
    
    \item \textbf{Improved Optimization Algorithms:} Although we employed a Genetic Algorithm (GA) for strategy optimization, exploring other optimization techniques, such as reinforcement learning or dynamic programming, could further enhance the efficiency and scalability of QuSplit.
    
    \item \textbf{Real-World Implementation:} Finally, deploying QuSplit in real quantum computing environments would provide valuable insights into its practical performance and challenges, such as communication overhead and hardware constraints.
\end{itemize}

In conclusion, the proposed framework QuSplit opens new avenues for optimizing quantum computing workflows from a job-splitting perspective in the era of NISQ devices.

\begin{acks}
This work was supported in part by the NSF 2311949. This work was also supported by the U.S. Department of Energy, Office of Science, Office of Advanced Scientific Computing Research through the Accelerated Research in Quantum Computing Program MACH-Q Project. The research used IBM Quantum resources via the Oak Ridge Leadership Computing Facility at the Oak Ridge National Laboratory, which is supported by the Office of Science of the U.S. Department of Energy under Contract No. DE-AC05-00OR22725.
\end{acks}

\bibliographystyle{ACM-Reference-Format}
\bibliography{bib}

\end{document}